\documentclass[12pt]{article}
\def\nc{\newcommand}
\nc{\bA}{\mbox{\boldmath $A$\unboldmath}}
\nc{\bn}{\mbox{\boldmath $n$\unboldmath}}
\nc{\bl}{\mbox{\boldmath $l$\unboldmath}}
\nc{\bm}{\mbox{\boldmath $m$\unboldmath}}
\def\nn{\nonumber}
\def\be{\begin{equation}}
\def\ee{\end{equation}}
\def\bea{\begin{eqnarray}}
\def\eea{\end{eqnarray}}

\nc{\cD}{\cal D}
\nc{\cL}{\cal L}
\nc{\cLd}{{\cal L}^{\dagger}}
\nc{\sta}{\sin\theta}
\nc{\cta}{\cos\theta}
\nc{\sda}{\sin^2\theta}
\nc{\cda}{\cos^2\theta}
\nc{\coa}{\cot\theta}
\nc{\sqd}{\sqrt{2}}

\nc{\p}{\partial}
\nc{\pr}{\p_r}
\nc{\pv}{\p_v}
\nc{\pta}{\p_{\theta}}
\nc{\pvi}{\p_{\varphi}}
\nc{\pdr}{\p^2_ r}
\nc{\pdv}{\p^2_v}
\nc{\pdta}{\p^2_{\theta}}
\nc{\pdvi}{\p^2_{\varphi}}
\nc{\pdvr}{\p^2_{vr}}
\nc{\pdra}{\p^2_{r\theta}}

\nc{\spr}{\frac{\p}{\p r_*}}
\nc{\spv}{\frac{\p}{\p v_*}}
\nc{\spta}{\frac{\p}{\p \theta_*}}
\nc{\spdr}{\frac{\p^2}{\p r_*^2}}
\nc{\spdvr}{\frac{\p^2}{\p r_* \p v_*}}
\nc{\spdra}{\frac{\p^2}{\p r_* \p \theta_*}}
\def\drH{\dot{r}_h}
\def\prH{r_h^{\prime}}
\def\pprH{r_h^{\prime\prime}}

\begin{document}
\baselineskip 18pt

\title{\bf\large Hawking Radiation of Weyl Neutrinos in a Rectilinearly
Non-uniformly Accelerating Kinnersley Black Hole\footnote{Supported by
the NNSF of China under Grant number: 19875019}}
\author{Wu Shuang-Qing 
\thanks{E-mail: sqwu@iopp.ccnu.edu.cn}
and Cai Xu 
\\
\it \small Institute of Particle Physics, Hua-Zhong Normal
University, Wuhan 430079, China}
\date{}
\maketitle

\begin{quote}
\hspace*{0.8cm}
Quantum thermal effect of Weyl neutrinos in a rectilinearly non-uniformly
accelerating Kinnersley black hole is investigated by using the generalized
tortoise coordinate transformation. The equation that determines the location,
the Hawking temperature of the event horizon and the thermal radiation
spectrum of neutrinos are derived. Our results show that the location and
the temperature of the event horizon depend not only on the time but also
on the angle.


{\bf PACC}: 0420, 9760L
\end{quote}

\vskip 1.5cm
\section{Introduction}
\hspace*{0.8cm}
The discovery of Hawking radiation$^{\cite{Hawk}}$ has not only solved
some contradictions inside black hole thermodynamics but also revealed a
profound intrinsic relations among quantum mechanics, thermodynamics and
general relativity. It is an important subject of black hole physics to
investigate the thermal properties of various black holes.$^{\cite{Zhao}}$
Thanks to the existence of various astrophysical processes such as evaporation
and accretion,$^{\cite{XM}}$ a black hole in the universe must change with time,
thus to make a deeper investigation on the Hawking radiation of an evaporating black
hole is of much significance to understand such dark celestial bodies as black
holes. In recent years, researches on the Hawking radiation of non-stationary
black holes attract much more attention, but most of these studies concentrated
on the quantum thermal effect of scalar fields. The investigation of the Hawking
radiation of fermions in the non-static black holes is merely limited to those
in spherically symmetric ones.$^{\cite{Zhao}}$ Recently, by making use of the
generalized tortoise transformation proposed by Zhao Zheng,$^{\cite{Zhao}}$ Wu
and Cai$^{\cite{WC1}}$ has successfully dealt with the Hawking radiation of Dirac
particles in a non-stationary Kerr black hole.$^{\cite{CM}}$ They simultaneously
treated the first-order and second-order Dirac equations and derived both the
event horizon equation and the Hawking temperature which coincide with previous
results.$^{\cite{NK}}$ A New quantum effect observed in the thermal radiation
spectrum of fermions probably originates from the coupling of the spin of particles
with the angular momentum of black holes. It is absent from the Bosonic radiation
spectrum of scalar particles.$^{\cite{NK}}$

In this paper, we use the method developed in Ref. \cite{WC1} to cope with the
Hawking radiation of massless fermions namely Weyl neutrinos in a rectilinearly
non-uniformly accelerating Kinnersley black hole.$^{\cite{Kinn}}$ The corresponding
quantum thermal effect of scalar particles in the same space-time has already been
studied in Ref. \cite{LZ}.

\section{The Weyl equation
in the background space-time}
\hspace*{0.8cm}
The metric of a rectilinearly non-uniformly accelerating Kinnersley
black hole that we will take into account can be rewritten in the
advanced Eddington-Finkelstein coordinate system as$^{\cite{Kinn}}$
\be
ds^2 = 2dv(G dv -dr -r^2f d\theta) -r^2(d\theta^2 +\sin^2\theta
d\varphi^2) \, ,
\ee
where $2G = 1 -2M/r -2a r\cta -r^2f^2$, $f = -a\sta$,
the parameter $a = a(v)$ is the magnitude of acceleration,
the mass $M(v)$ of the hole is a function of time $v$.

We choose such a complex null-tetrad
$\{\bl, \bn, \bm, \overline{\bm}\}$
that satisfies the orthogonal conditions
$\bl \cdot \bn = -\bm \cdot \overline{\bm} = 1$.
Thus the covariant one-forms can be written as
\bea
&&\bl = dv \, ,~~~~~~\bn = G dv -dr -r^2f d\theta  \, , \nn\\
&&\bm = \frac{-r}{\sqd}\left(d\theta +i\sta d\varphi\right) \, , \nn\\
&&\overline{\bm} = \frac{-r}{\sqd}\left(d\theta -i\sta d\varphi\right) \, ,
\eea
and their corresponding directional derivatives are
\bea
&&D = -\pr \, , ~~~~~~\Delta = \pv +G\pr \, , \nn \\
&&\delta = \frac{1}{\sqd r}\left(-r^2f \pr +\pta
+\frac{i}{\sta}\pvi\right) \, , \nn\\
&&\overline{\delta} = \frac{1}{\sqd r}\left(-r^2f \pr
+\pta -\frac{i}{\sta}\pvi\right) \, .
\eea
It is not difficult to compute the non-vanishing Newman-Penrose
(NP)$^{\cite{NP}}$ spin coefficients as follows (we denote $G_{,r} = dG/dr$)
\bea
&&\rho = 1/r \, , ~~\mu = G/r \, , ~~\gamma = -G_{,r}/2 \, ,
~~\tau = -\tilde{\pi} = f/\sqd \, , ~~\beta = \coa/(2\sqd r) \, ,\nn\\
&&\alpha = -\coa/(2\sqd r) +f/\sqd\, ,
~~\nu = \left[(2rG -r^2G_{,r})f +r^2f_{,v} +G_{,\theta} \right]/(\sqd r) \, .
\eea

Inserting the related spin coefficients and directional derivatives
into the spinor form of the Weyl neutrino equation$^{\cite{CM,Teuk}}$
in the NP formalism
\bea
&&(D +\epsilon -\rho)\eta_1 -(\overline{\delta} +\tilde{\pi}
-\alpha)\eta_0 = 0 \, , \nn\\
&&(\Delta +\mu -\gamma)\eta_0
-(\delta +\beta -\tau)\eta_1 = 0 \, ,
\eea
we obtain
\bea
&&{\cD}_1 \eta_1 +\frac{1}{\sqd r}\left({\cL}
-r^2f{\cD}_2\right) \eta_0 = 0 \, , \nn\\
&&\left(\pv +G{\cD}_1 +G_{,r}/2\right) \eta_0
-\frac{1}{\sqd r}\left({\cLd} -r^2f{\cD}_1\right) \eta_1 = 0 \, ,
\label{weyl}
\eea
in which we have defined operators
$${\cD}_n = \pr +n/r \,
, ~~{\cL} = \pta +\frac{1}{2}\coa -\frac{i}{\sta}\pvi  \, ,
~~{\cLd} = \pta +\frac{1}{2}\coa +\frac{i}{\sta}\pvi  \, .$$

With further substitutions $\chi_1 = \sqd r \eta_1$, $\chi_0 = \eta_0$,
into Eq. (\ref{weyl}), we can reduce them to the form
\bea
&&{\cD}_0 \chi_1 +\left({\cL}
-r^2f{\cD}_2\right) \chi_0 =  0 \, , \nn\\ &&r^2\left(2\pv
+2G{\cD}_1 +G_{,r}\right) \chi_0 -\left({\cLd} -r^2f{\cD}_0\right)
\chi_1 = 0  \, . \label{fowe}
\eea

\section{Event
horizon equation}
\hspace*{0.8cm}
Though Eq. (\ref{fowe}) can not be decoupled with respect to
$\theta$, however to deal with the problem of Hawking radiation,
one should be concerned about its asymptotic behaviors near the
horizon only. Basing upon of the symmetry of the space-time under
consideration, we introduce a generalized tortoise coordinate
transformation as follows
\be
r_* = r +\frac{1}{2\kappa}\ln[r -r_h(v,\theta)] \, , ~~v_* = v
-v_0 \, , ~~\theta_* = \theta -\theta_0 \, , \label{trans}
\ee
namely,
$$ dr_* = dr +\frac{1}{2\kappa(r -r_h)} \left(dr -\drH dv -\prH
d\theta \right) \, , ~~dv_* = dv \, , ~~d\theta_* = d\theta \, ,
$$
where $r_h = r_h(v,\theta)$ is the location of the event horizon.
$\kappa$ is an adjustable parameter to be determined, the parameters
$v_0$ and $\theta_0$ are arbitrary constants, all of them remain
unchanged under the tortoise coordinate transformation. $\drH =
\pv r_h$ and $\prH = \pta r_h$ can be viewed as parameters
depicting the evolution of the event horizon. All induced
relations among the first-order and second-order derivatives
from the transformation (\ref{trans}) can be referred to
Ref. \cite{Zhao} and are not listed here.

Now we apply the transformation (\ref{trans}) to Eq. (\ref{fowe})
and take the $r \rightarrow r_h(v_0,\theta_0)$, $v \rightarrow v_0$
and $\theta \rightarrow \theta_0$ limits, then we can obtain its
limiting form near the event horizon as
\bea
&&\spr \chi_1 -\left(\prH +r_h^2f \right)\spr \chi_0 = 0 \, , \nn\\
&&\left(\prH +r_h^2f\right)\spr \chi_1
+2r_h^2\left(G -\drH\right)\spr \chi_0 = 0 \, .
\label{rela}
\eea

Obviously the condition that Eq. (\ref{rela}) has a nontrial
solution for $\chi_0$ and $\chi_1$ is that its determinant vanishes,
which gives the following equation to determine the location
of the event horizon
\be
2G -2\drH +\Big(r_hf +\frac{\prH}{r_h}\Big)^2 = 0 \, .
\label{loca}
\ee
It is easily to see that this result is  consistent with that
inferred from the null-surface condition: $g^{ij}\p_i F\p_j F = 0$.
Making a similar prescription as done in the above to the null-surface
equation
\be
(2G +r^2f^2)(\pr F)^2  +2\pv F\pr F -2f\pta F\pr F
+\frac{1}{r^2}(\pta F)^2 +\frac{1}{r^2\sda} (\pvi F)^2 = 0 \, ,
\ee
we can arrive at an equation
$$\left(2G +r_h^2f^2 -2\drH +2f\prH
+\frac{{\prH}^2}{r_h^2}\right)\left(\spr F\right)^2 = 0 \,  ,$$
which results in the same one (\ref{loca}) that determines
the event horizon after we let the coefficient in the front of the
derivative $\left(\spr F\right)^2$ to be zero. Eq. (\ref{loca})
shows that $r_h$ depends not only on $v$ but also on $\theta$.
This means that the location of the event horizon and the shape
of the black hole change with time.

\section{Hawking temperature}
\hspace*{0.8cm}
To investigate the Hawking radiation of Weyl neutrinos, we are now
in a position to consider the second-order form of the Weyl equation.
After a direct calculation to the second-order Weyl equations
\be
[r^2{\cD}_2(2\pv +2G{\cD}_1 +G_{,r}) +({\cLd} -r^2f{\cD}_2)({\cL}
-r^2f{\cD}_2)] \chi_0 = 0 \, ,
\ee
and
\bea
&&[r^2(2\pv +2G{\cD}_1 +G_{,r}){\cD}_0
+({\cL} -r^2f{\cD}_0)({\cLd} -r^2f{\cD}_0)] \chi_1 \nn \\
&& = 2r^2\Big\{[(2rG -r^2G_{,r})f+r^2f_{,v}+G_{,\theta}]\pr \nn\\
&&~~+\left(rG_{,r} -3G +r^2G_{,rr}/2\right)f +2rf_{,v} +G_{,\theta}/r
+G_{,r\theta}/2\Big\}\chi_0  \, ,
\eea
we can get their explicit forms as follows
\bea
&&\Big[(2r^2G +r^4f^2)\pdr +2r^2\pdvr +\pdta +\frac{1}{\sda}\pdvi
 -2r^2f\pdra +\frac{i\cta}{\sda}\pvi\nn\\
&&+(\coa -4rf)\pta  +4r\pv +(3r^2G_{,r} +6rG +6r^3f^2 -2r^2f\coa)\pr \nn\\
&&+r^2G_{,rr} +4rG_{,r} +2G +6r^2f^2 -4rf\coa
-\frac{1}{4\sda} -\frac{1}{4} \Big]\chi_0 = 0 \, ,
\label{soe1}
\eea
and
\bea
&&\Big[(2r^2G +r^4f^2)\pdr +2r^2\pdvr +\pdta +\frac{1}{\sda}\pdvi
-2r^2f\pdra -\frac{i\cta}{\sda}\pvi \nn\\
&&+\coa \pta+(r^2G_{,r} +2rG +2r^3f^2 -2r^2f\coa)\pr
-\frac{1}{4\sda} -\frac{1}{4} \Big]\chi_1 \nn\\
&& = 2r^2\Big\{[(2rG -r^2G_{,r})f+r^2f_{,v}+G_{,\theta}]\pr \nn\\
&&~~+\left(rG_{,r} -3G +r^2G_{,rr}/2\right)f +2rf_{,v} +G_{,\theta}/r
+G_{,r\theta}/2\Big\}\chi_0  \, .
\label{soe2}
\eea

Application of a similar procedure as in the last section to
Eqs. (\ref{soe1},\ref{soe2}) leads to
\bea
&&\left[\frac{A}{2\kappa} +2r_h^2(2G -\drH) +2r_h^4f^2
+2fr_h^2\prH\right] \spdr\chi_0 -2\left(f r_h^2 +\prH\right) \spdra\chi_0  \nn\\
&& +2r_h^2 \spdvr\chi_0  +\Big[-A +3r_h^2G_{,r} +r_h(6G -4\drH) +6r_h^3f^2 \nn\\
&& -2r_h^2f\coa_0 +(4f r_h -\coa_0)\prH  -\pprH \Big] \spr\chi_0 = 0 \, ,
\label{sowe1}
\eea
and
\bea
&&\left[\frac{A}{2\kappa} +2r_h^2(2G -\drH) +2r_h^4f^2
+2fr_h^2\prH\right] \spdr\chi_1 -2\left(f r_h^2 +\prH\right) \spdra \chi_1 \nn\\
&&+2r_h^2 \spdvr\chi_1 +\Big[-A +r_h^2G_{,r} +2r_hG +2r_h^3f^2 -2r_h^2f\coa_0
-\coa_0\prH  \nn\\
&&-\pprH \Big] \spr\chi_1 = 2r_h^2\left[(2r_hG -r_h^2G_{,r})f +r_h^2f_{,v}
+G_{,\theta}\right] \spr\chi_0  \nn\\
&&= -\frac{\prH +r_h^2f}{G -\drH}\left[(2r_hG -r_h^2G_{,r})f
 +r_h^2f_{,v} +G_{,\theta}\right] \spr \chi_1 \, ,
\label{sowe2}
\eea
in which the coefficient $A$ is an infinite form of $0/0$-type. Its value
can be computed by means of L' H\^{o}spital rule as
\bea
A &=& \lim_{r \rightarrow r_h(v_0,\theta_0)}
\frac{2r^2(G -\drH) +r^4f^2 +2f r^2\prH +{\prH}^2}{r -r_h} \nn\\
&=& 2r_h^2G_{,r} +4r_h(G -\drH) +4r_h^3f^2 +4f r_h\prH \nn\\
&=& 2r_h^2G_{,r} +2r_h^3f^2 -2{\prH}^2/r_h \, . \nn
\eea

Now we adjust the parameter $\kappa$ in Eqs. (\ref{sowe1},\ref{sowe2})
such that the ratio of the coefficients before $\spdr$ and $\spdvr$ is
equal to $1 : 2$, namely
$$\frac{A}{2\kappa} +2r_h^2(2G -\drH) +2r_h^4f^2
+2f r_h^2\prH \equiv r_h^2 \, , $$
and get the surface gravity of the event horizon
\be
\kappa =\frac{r_h^2G_{,r} +r_h^3f^2
-{\prH}^2/r_h}{r_h^2(1 -2G) -r_h^4f^2 +{\prH}^2} \, .
\ee

Dividing both sides of Eqs. (\ref{sowe1},\ref{sowe2})by $r_h^2$
after substituting the coefficient $A$ into them, we can reduce
both of them to a standard wave equation near the event horizon
\be
\spdr \Psi +2\spdvr \Psi -2B \spdra \Psi + 2C \spr \Psi = 0 \, ,
\label{wave}
\ee
where $B = f +\prH/r_h^2$, for $\Psi = \chi_0$ we have
$$2C = \frac{2G}{r_h} +G_{,r} +2r_hf^2 -(2f
r_h +\frac{\prH}{r_h^2})\coa_0 -\frac{\pprH}{r_h^2} \, ,$$
while for $\Psi = \chi_1$ we have
\bea
2C &=& \frac{2G}{r_h} -G_{,r} -(2f
+\frac{\prH}{r_h^2})\coa_0 +\frac{2{\prH}^2}{r_h^3}
-\frac{\pprH}{r_h^2} \nn\\
 &+& \frac{f+\prH/r_h^2}{G-\drH}\left[(2r_hG -r_h^2G_{,r})f +r_h^2f_{,v}
+G_{,\theta}\right] \, . \nn
\eea

From the above prescriptions, we can see that it is with the aid
of the relations (\ref{rela}) derived in the preceding section
that we can recast the coupled equation (\ref{sowe2}) into a
second-order wave equation of one component. This demonstrates
that our disposition of the first-order and second-order equations
is self-consistent and meets with the needs on physical grounds.

\section{Thermal radiation spectrum}
\hspace*{0.8cm}
Now that all coefficients in Eq. (\ref{wave}) can be regarded as finite
real constants, one can separate variables as
\be
\Psi = R(r_*)\Theta(\theta_*)e^{i(m\varphi-\omega v_*)} \, ,
\label{seva}
\ee
and with this ansatz substituted into equation
(\ref{wave}), one gets the following equation
$$\Theta^{\prime} =
\lambda \Theta \, , ~~~R^{\prime\prime} = 2(i\omega -C_0)
R^{\prime} \, ,$$
with their solutions being
\be
\Theta = e^{\lambda \theta_*} \, ,
~~~~R = C_1e^{2(i\omega -C_0)r_*} +C_2 \, ,
\ee
where $\lambda$ is a real constant introduced in the separation variables,
$C_0 = C -\lambda B$.

The in-going wave and the out-going wave are, respectively
\bea
\Psi_{\rm in} &\sim& e^{i(m\varphi -\omega v_*) +\lambda \theta_*}
\, , \nn\\
\Psi_{\rm out} &\sim& e^{i(m\varphi -\omega v_*)
+\lambda \theta_*}e^{2(i\omega -C_0)r_*} \nn \\
&=& \Psi_{\rm in}e^{2(i\omega -C_0)r_*} \, ,~~~~(r > r_h) \, .
\eea
Near the event horizon, we have $r_* \sim \frac{1}{2\kappa}\ln (r - r_h)$.
The in-going wave $\Psi_{\rm in}$ is regular at the event horizon,
but the out-going wave $\Psi_{\rm out}(r > r_h)$ is irregular at
the event horizon $r=r_h$, it can be analytically extended from
the outside of the hole into the inside of the hole through the
lower complex $r$-plane
$$ (r -r_h) \rightarrow (r_h -r)e^{-i\pi}$$ to
\bea
\widetilde{\Psi_{\rm out}} &=& e^{i(m\varphi
-\omega v_*) +\lambda \theta_*} e^{2(i\omega
-C_0)r_*}e^{\pi(\omega +iC_0)/\kappa} \nn\\
 &=& \Psi_{\rm out}e^{\pi(\omega +iC_0)/\kappa} \, , ~~~~ (r < r_h) \, .
\eea

According to the method suggested by Damour-Ruffini-Sannan,$^{\cite{DRS}}$
the relative scattering probability of the outgoing wave at the horizon is
\be
\left|\frac{{\Psi}_{\rm out}}{\widetilde{\Psi_{\rm out}}}\right|^2
= e^{-2\pi\omega/\kappa} \, ,
\ee
and the thermal radiation Fermionic spectrum of Weyl neutrinos from the event
horizon of the hole is given by
\be
\langle {\cal N}(\omega) \rangle = \frac{1}{e^{\omega/T_h } + 1} \, ,
\label{sptr}
\ee
with the obvious expression of the Hawking temperature being
\be
T_h = \frac{\kappa}{2\pi} = \frac{1}{4\pi r_h} \cdot \frac{M r_h
-r_h^3a\cta_0 -{\prH}^2}{M r_h +r_h^3a\cta_0 +{\prH}^2/2} \, .
\label{temp}
\ee
It follows that the temperature depends not only on the time, but also on
the angle $\theta$. The temperature is in accord with the corresponding result
presented in Ref. \cite{LZ}.

\section{Conclusions}
\hspace*{0.8cm}
Equations (\ref{loca}) and (\ref{temp}) give the location and
the temperature of event horizon, which depend not only on the
advanced time $v$ but also on the polar angle $\theta$. They are
just the same results as that obtained in the discussion of the
thermal radiation of scalar fields.$^{\cite{LZ}}$ Equation (\ref{sptr})
shows the Hawking radiation spectrum of Weyl neutrinos in a
rectilinearly non-uniformly accelerating Kinnersley black hole.

This study and those in Refs. [4,12-14] manifest that the method
of generalized coordinate transformation is a powerful tool to
investigate black hole radiation. It was initially proposed by
Damour-Ruffini$^{\cite{DRS}}$ to deal with the quantum thermal
effect of scalar fields in a static (Schwarzshild) black hole
and in a stationary (Kerr) black hole. Later Sannan$^{\cite{DRS}}$
developed this method and made it suitable not only to dealing
with the thermal radiation of scalar fields but also to that of
fermions. Zhao Zheng$^{\cite{Zhao}}$ further extended this transformation
in the static and stationary cases to that in the non-static and
non-stationary cases and have made several discussions about the
quantum thermal effects of scalar fields in various black holes
and that of Dirac particles in some non-static and spherically
symmetric black holes. But this method meets great difficulties
when it is applied to the Hawking radiation of Dirac particles in
a most general space-time such as a non-static and non-spherically
symmetric black hole or a non-stationary axisymmetry black
hole. The reason is that the Dirac equation can not be completely
separable for all variables in the most general space-time. Now
this difficulty is overcome when we simultaneously treat the
first-order and second-order equations, and use the relations
between the first-order derivatives to remove the crossing-terms
in the second-order equations. This procedure enables us to reduce
each second-order equation of one component to a simple separable
standard wave equation near the event horizon after we have made a
generalized tortoise coordinate transformation.

In conclusion, the work in this paper and those in Refs. [4,12-14] make
the method of generalized tortoise transformation more perfect. They
indicate that this theory becomes a fairly integrated system. The method
developed by us is applicable not only to discussing the Hawking
radiation of Weyl neutrinos in a non-static and non-spherically
symmetric Kinnersley black hole with arbitrarily acceleration and
that of electrons in a non-stationary axisymmetry black hole, but
also, in principle, to studying the quantum thermal effect of any
space-time with a non-degenerate event horizon.$^{\cite{WC4}}$ Also
it is easily extended to discuss the Hawking effect of fields with
an arbitrary spin.$^{\cite{WC2,WC3}}$

\end{document}